\documentclass[twocolumn]{aastex6}
\usepackage{rotating}
\usepackage{graphicx}
\usepackage{amssymb}
\usepackage{amsmath}
\usepackage{hyperref}

\def\cxo{{\sl CXO\ }}

\def\xmm{{\sl XMM-Newton\ }}


\def\gtrsim{\mathrel{\hbox{\rlap{\hbox{\lower4pt\hbox{$\sim$}}}\hbox{$>$}}}}
\def\lesssim{\mathrel{\hbox{\rlap{\hbox{\lower4pt\hbox{$\sim$}}}\hbox{$<$}}}}
\def\farcs{\hbox{$.\!\!^{\prime\prime}$}}

\begin{document}

\title{XMM-Newton and Chandra observations of the unidentified Fermi-LAT source 3FGL J1016.5--6034: A Young Pulsar with a Nebula?}
\author{Jeremy Hare\altaffilmark{1,2}, Igor Volkov\altaffilmark{2}, Oleg Kargaltsev\altaffilmark{2}, George Younes\altaffilmark{2}, Blagoy Rangelov\altaffilmark{3}}
\altaffiltext{1}{Space Sciences Laboratory, 7 Gauss Way, University of California, Berkeley, CA 94720, USA}
\altaffiltext{2}{Department of Physics, The George Washington University, 725 21st St. NW, Washington, DC 20052, USA}
\altaffiltext{3}{Texas State University, Department of Physics, 601 University Drive, San Marcos, TX 78666, USA}
\email{jeh86@gwu.edu}

\begin{abstract}

We report the discovery of a 
  bright X-ray source in the  {\sl XMM-Newton} and {\sl Chandra X-ray Observatory} ({\sl CXO}) images  of the unidentified {\sl Fermi-}LAT source 3FGL J1016.5--6034. 
    The {\sl XMM-Newton} spectrum of the source is well fit by an absorbed blackbody+power-law model with a temperature, $kT=0.20\pm0.02$ keV, and photon index $\Gamma=1.8\pm0.1$. The {\sl CXO}   resolves the same  source into a point source (CXOU~J101546.0--602939) and a surrounding compact nebula seen up to about $30''$ from the point source. The {\sl CXO}  spectrum of the nebula can be described by an absorbed power-law  with $\Gamma=1.7\pm0.3$ and is partly responsible for the non-thermal emission observed in the  {\sl XMM-Newton}  spectrum.  The {\sl XMM-Newton}  images also reveal faint extended emission on arcminute scales. These properties strongly suggest that the X-ray source and the accompanying  extended emission are a newly discovered young pulsar with a pulsar wind nebula.  We also analyze $\sim10$ years of {\sl Fermi-}LAT data and find that the improved LAT source localization is consistent with the position of CXOU~J101546.0--602939. 
\end{abstract}

\section{Introduction}
Prior to the launch of the {\sl Fermi Gamma-ray Space Telescope} (hereafter {\sl Fermi}) only seven gamma-ray pulsars were known. Over the past 10 years this number has grown to over 200\footnote{See e.g., \url{https://confluence.slac.stanford.edu/display/GLAMCOG/Public+List+of+LAT-Detected+Gamma-Ray+Pulsars}}, thanks to the {\sl Fermi} Large Area Telescope (LAT). The increase in sensitivity at GeV energies provided by the {\sl Fermi-}LAT  has allowed for a fainter population of new $\gamma$-ray pulsars to be discovered at GeV energies (see e.g., \citealt{2017ApJ...834..106C}). However, it is often difficult to detect a period in an unidentified GeV source hosting a GeV pulsar in a ``blind'' search because one needs to search for both the period and period derivative (see e.g., \citealt{2010ApJ...725..571S,2018arXiv181200719S}), which is computationally expensive.
  Therefore, it is often beneficial to search for a lower-energy counterpart (e.g., radio, X-rays) to an unidentified {\sl Fermi} source suspected to be a GeV pulsar in order to understand the nature of the $\gamma$-ray source.  Since a large fraction of Galactic GeV source are pulsars,  sensitive X-ray observations can be used to search for X-ray pulsations or for diffuse emission from  a pulsar-wind  nebula (PWN) associated with the pulsar.  
  
  We  selected the unidentified source 3FGL J1016.5--6034 (hereafter  J1016) from the  Third {\sl Fermi}-LAT Source Catalog   (3FGL; \citealt{2015ApJS..218...23A})  to study in X-rays with the {\sl Chandra X-ray Observatory} ({\sl CXO}) and the {\sl X-ray Multi-Mirror Mission}  ({\sl XMM-Newton}). According to the 3FGL catalog  J1016  exhibits a pulsar-like  spectrum at GeV energies (i.e., having significant spectral curvature and a soft photon index $\gamma>2.5$)  with a flux of $f_{\rm GeV}=2\times10^{-11}$ erg s$^{-1}$ cm$^{-2}$ in 0.1-100 GeV and lies relatively close to the Galactic plane  ($l=285.0^{\circ}$, $b=-3.2^{\circ}$).  
  
  Here we discuss \cxo and \xmm observations of  J1016. In Section \ref{sec2} we describe the observations and data reduction, in Sections \ref{sec3} the {\sl CXO}, {\sl XMM-Newton}, and {\sl Fermi-}LAT
 data analysis,  followed by the discussion in Section \ref{sec4} and summary of the results in Section \ref{sec5}.




\section{Observations and Data Reduction}
\subsection{{\sl XMM-Newton}}
\label{sec2}
The field of 3FGL J1016 was observed by {\sl XMM-Newton} on 2017 May 25 for 18 ks (obsID 0802930101). The EPIC pn and both MOS detectors were operated in Full Frame mode,  offering time resolutions of 73.4 ms and 2.6 s, respectively. We reduced and analyzed the {\sl XMM-Newton} data using the Science Analysis System (SAS) version 16.1.0. Event lists for the pn, MOS1, and MOS2, were cleaned (e.g., removing times with high particle background) and calibrated following standard SAS procedures. After cleaning, scientific exposures of 12 ks, 16.4 ks, and 16.4 ks, remained for the pn, MOS1, and MOS2, respectively.   The total observation time span is  15.0 ks, 16.6 ks, and 16.6 ks in   pn, MOS1, and MOS2, respectively.

\begin{figure*}
\centering
\includegraphics[trim={0 0 0 0},scale=0.6]{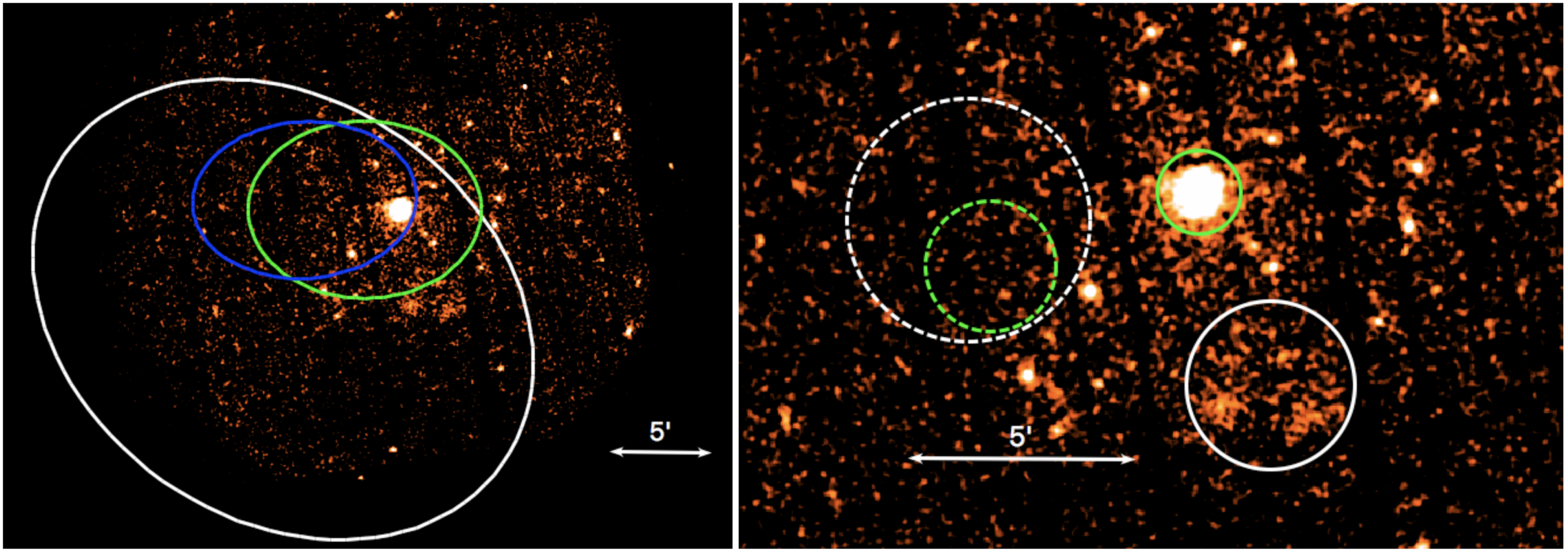}
\caption{{\sl Left:} {\sl XMM-Newton} binned and smoothed EPIC mosaic image of the J1016 field.  The white ellipse shows the 95\% positional uncertainty ellipse of  J1016 from the 3FGL catalog \citep{2015ApJS..218...23A}. The green ellipse shows the newly derived GeV source position and systematic plus statistical uncertainties from 10 years of {\sl Fermi-}LAT data using a PL model (see Section \ref{lat_ana_specim}). For completeness, we include the source GeV source position derived using a LP model, as it slightly differs from the PL position. CXOU J101546 is contained in all three error ellipses and is by far  the brightest source in the field. 
  {\sl Right:} Zoomed-in image of the CXOU J101546 vicinity. 
Faint extended emission can be seen about $5'$  to the south of the bright X-ray source. The green regions show the source (solid) and background (dashed) regions used for the point source spectral extraction. The white circles show the source (solid) and background (dashed)  regions used for the extended emission spectral extraction (see Section \ref{xspec}). East is to the left and North is up in both images.
\label{xmm}
}
\end{figure*}

The SAS task {\tt edetect\_chain} was then used to detect X-ray sources  in the five standard energy bands used by the 3XMM-DR8 catalog\footnote{ See e.g., \url{http://xmmssc.irap.omp.eu/Catalogue/3XMM-DR8/3XMM-DR8_Catalogue_User_Guide.html}} \citep{2016A&A...590A...1R}. In total, 54 X-ray sources were detected in the field of 3FGL J1016 with a likelihood $\geq10$, corresponding to a 4$\sigma$ detection significance\footnote{ See \url{https://xmm-tools.cosmos.esa.int/external/sas/current/doc/emldetect.pdf}}. Of these 54 sources, one source 
 stands out 
  because it has a factor of $\sim$20 more counts than the next brightest X-ray source and is the only source that has enough photons to fit a constraining spectrum.
   Therefore, we extracted the spectrum of this source from all three EPIC detectors using a circular region centered on the source position with a 55$''$ radius. A source-free region offset from the source was chosen for the background region (see Figure \ref{xmm}). Prior to fitting we binned the spectrum to contain a minimum of 25 counts per bin.

All X-ray spectra in this paper are fit using XSPEC (version 12.10.1; \citealt{1996ASPC..101...17A}). The Tuebingen-Boulder ISM absorption model ({\tt tbabs}) with the solar abundances of \cite{2000ApJ...542..914W} were used. The uncertainties reported in this paper are all 1 $\sigma$ unless otherwise noted.

\subsection{Chandra X-ray Observatory}
\label{cxo_ana}
J1016 was also observed with the {\sl Chandra} Advanced CCD Imaging Spectrometer (ACIS) 
on 2018 January 15 for $\sim2$ ks (obsID 20090).  The 
 source was imaged on the ACIS-I detector operated in timed exposure mode using the Very Faint telemetry format, which offers a time resolution of 3.2 s. The {\sl Chandra} data were processed and analyzed using the {\sl Chandra} Interactive Analysis of Observations (CIAO) package version 4.11 with the Calibration Database (CALDB) version 4.8.2. The data were processed and cleaned using standard CIAO procedures. 

\begin{figure*}
\centering
\includegraphics[trim={0 0 0 0},scale=0.6]{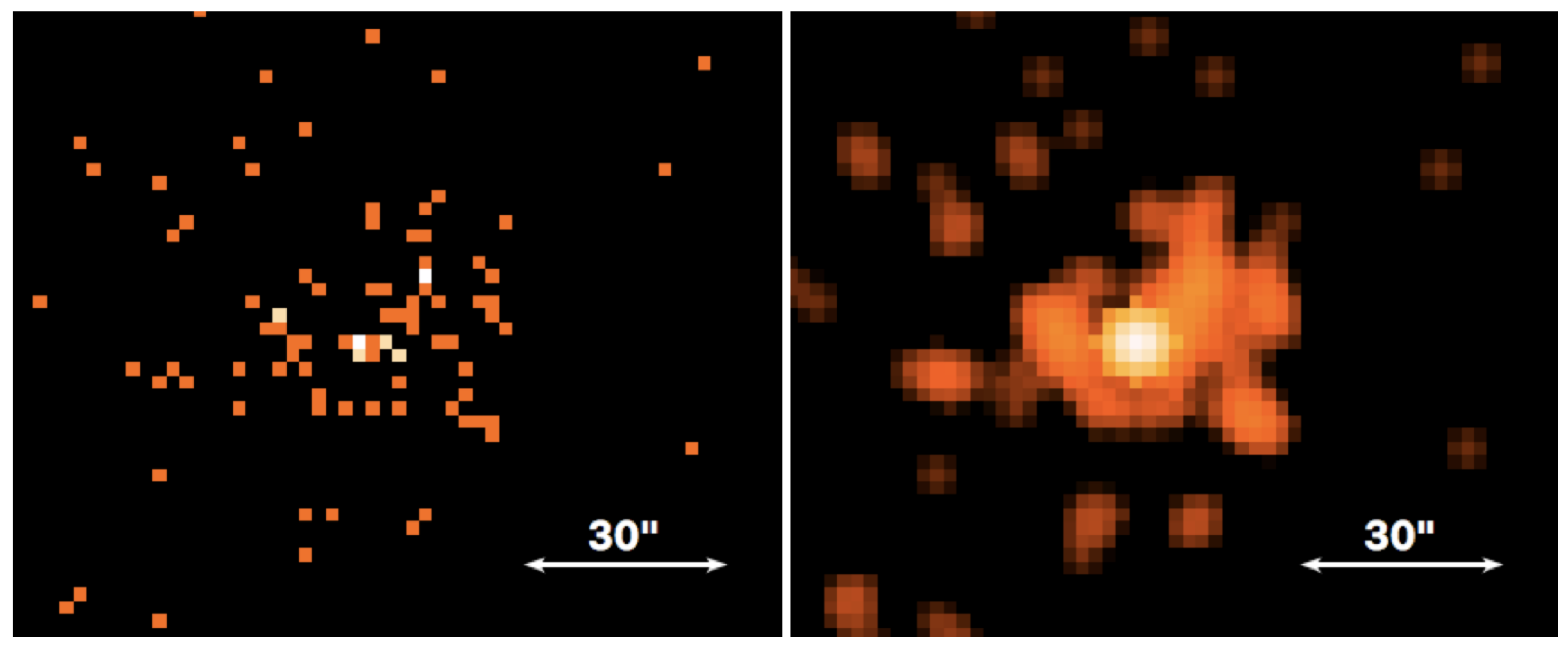}
\caption{The left panel shows the binned (by a factor of 4) {\sl CXO} ACIS-I zoomed in image of the brightest X-ray source in the field.  The source is resolved by {\sl CXO} into a point source and surrounding diffuse emission. The right panel is the same as the left panel, but is smoothed with a 3$''$ gaussian kernel. East is to the left and North is up.
\label{cxo}
}
\end{figure*}

We used CIAO's {\tt wavdetect} task  with wavelet scales of 1, 2, 4, 8, and 16 pixels to detect X-ray sources  in the 0.5-7 keV energy range. Only one source (CXOU J101546.0-602939; CXOU J101546, hereafter),  which is surrounded by diffuse emission (see  Figure \ref{cxo}), is significantly detected in this observation due to its short duration. The source is detected at a significance of $\sim6\sigma$, even if we conservatively assume the surrounding diffuse emission as the local background. This source is coincident with the brightest source detected by {\sl XMM-Newton} and is located at a position R.A.$=153.94163(3)^{\circ}$ decl.$=-60.49431(2)^{\circ}$. Unfortunately, no other sources are confidently detected so we are unable to correct for any systematic uncertainty in the absolute astrometry of {\sl Chandra}. Therefore, the positional uncertainty is mostly dominated by the  {\sl Chandra} systematic uncertainty which we take to be  $0\farcs6$ at 68\% confidence or $0\farcs9$ at 95\% confidence\footnote{http://cxc.harvard.edu/cal/ASPECT/celmon/}. After adding the statistical and systematic uncertainties in quadrature, we obtain a 2$\sigma$ positional uncertainty of 0\farcs98. 

The CXOU~J101546 spectrum  was extracted from a $1\farcs5$ radius circle centered on the source's position. 
 The background was taken from an annulus with inner and outer radii, $r_{\rm in}=5''$ and $r_{\rm out}=25''$, which encompasses the extended emission surrounding the point source (see  Figure \ref{cxo}). For the nebula  spectrum, we chose an annulus centered on the point source with inner and outer radii $r_{\rm in}=5''$ and $r_{\rm out}=30''$. The background was taken from a source free circular region ($r=73''$) offset from the extended emission (centered on R.A.=$153.998^{\circ}$, decl.=$-60.465^{\circ}$). Given the small number of net counts, 33 and 57 for the point source and extended emission, respectively,  we used W-statistics\footnote{see  \url{https://heasarc.gsfc.nasa.gov/xanadu/xspec/manual/XSappendixStatistics.html} }, which is a variation of Cash statistics \citep{1979ApJ...228..939C} to fit their spectra.  
 
\subsection{Fermi-LAT}
\label{LATdata}
We downloaded the {\sl Fermi-}LAT data of the field of J1016, spanning  $\sim125$ months, observed between 2008 August 8  and 2019 January 13. The data were extracted from a 15$^{\circ}$ radius centered on CXOU J101546.0-602939 (i.e., R.A.=$153.942^{\circ}$, decl.=$-60.4943^{\circ}$) and were filtered to remove events with reconstructed zenith angles $<90^{\circ}$, in order to avoid events coming from the Earth limb. Our analysis includes data in the 100 MeV to 300 GeV energy range. The Pass 8 instrument response function (IRF) {\tt P8R3\_SOURCE\_V2} \citep{2013arXiv1303.3514A} was used and events were filtered such that they belonged to the ``Source'' event  class (i.e., evclass=128). The latest version (1.0.0) of the Fermitools\footnote{\url{https://github.com/fermi-lat/Fermitools-conda/wiki}} were used in conjunction with the analysis scripts provided by the Fermipy\footnote{\url{https://fermipy.readthedocs.io/en/latest/index.html}} python package \citep{2017ICRC...35..824W}.

The initial input models for J1016 and nearby sources were adopted from the 3FGL source catalog \citep{2015ApJS..218...23A}, while the Galactic diffuse emission and isotropic emission were modeled using {\tt gll\_iem\_v06} and {\tt iso\_P8R3\_SOURCE\_V2}, respectively\footnote{\url{https://fermi.gsfc.nasa.gov/ssc/data/access/lat/BackgroundModels.html}} \citep{2016ApJS..223...26A}. We include all sources within a 15$^{\circ}$ radius of 3FGL J1016 in our model. In the 3FGL catalog, J1016 is located at an R.A.=$154.135^{\circ}$ and decl.=$-60.576^{\circ}$ (the 95\% positional uncertainty ellipse is shown in Fig.~\ref{xmm})   
 and its spectrum is  best described  by a LogParabola (LP) spectral model \cite{2015ApJS..218...23A}.

\section{Results}
\label{sec3}
\subsection{X-ray data analysis}

Below we describe the results of  the image, spectral, and timing analyses of the {\sl XMM-Newton}  and \cxo observations.  These observations were part of the same observing program where the goal of the {\sl XMM-Newton} observation was to collect enough counts for spectral analysis and look for large-scale extended emission while the goal of the short \cxo observation was to  obtain an accurate position of the point source and to search for compact nebula around it. 

\subsubsection{Image Analysis}
\label{img_ana}
\begin{figure}
\centering
\includegraphics[trim={0 0 0 0},scale=0.3]{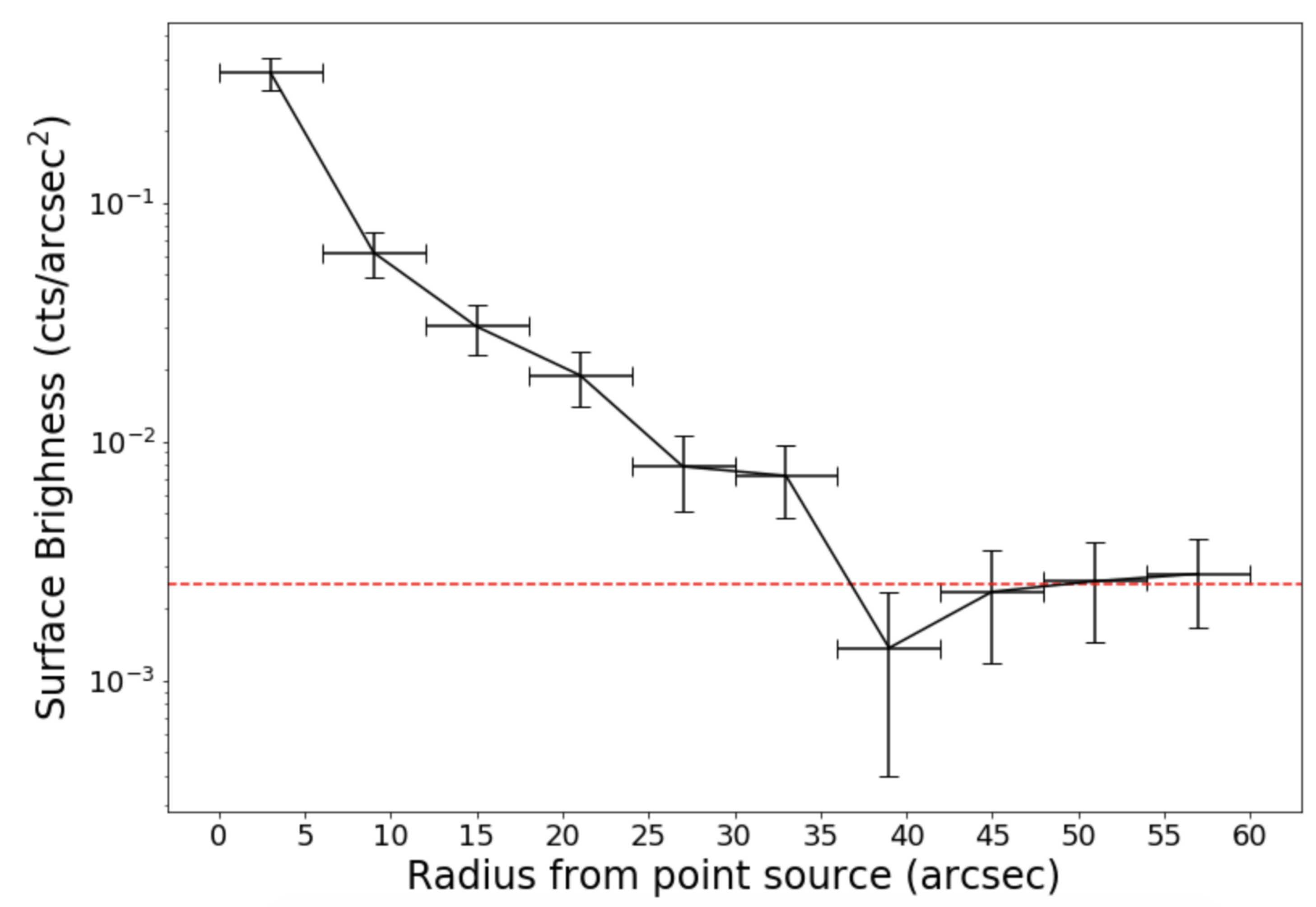}
\caption{ Radial profile of the extended emission seen in the {\sl Chandra} image shown in Figure \ref{cxo}. The red line shows the surface brightness of the background region used when extracting the energy spectrum of the extended nebula (see Section \ref{cxo_ana}). 
\label{rad_prof}
}
\end{figure}

Figure \ref{xmm} shows the 
  {\sl XMM-Newton} EPIC mosaic image of the J1016 field. The bright X-ray source  is located towards the edge of the 3FGL 95\% positional uncertainty ellipse of the {\sl Fermi}-LAT source (shown in the left panel of  Figure \ref{xmm}).  In addition to a number of faint unresolved sources, there is evidence of faint large scale emission located $\sim5'$ south of the bright X-ray source (shown by the solid white circle in the right panel of Figure \ref{xmm}). 
Figure \ref{cxo} shows the vicinity  of CXOU~J101546 as seen by {\sl Chandra}. 
The sub-arcsecond angular resolution of the {\sl CXO} resolves the bright X-ray source into a point source surrounded by compact diffuse emission seen out to a distance of $\sim 30''$.  The radial profile of the extended emission is shown in Figure \ref{rad_prof}.  The large-scale extended emission is too faint to be detected in the short  {\sl CXO} exposure. A deeper   {\sl CXO}  observation is needed to study the properties of  both the compact and large-scale extended emission.
  
\subsubsection{Spectral Analysis}
\label{xspec}
\begin{figure}
\centering
\includegraphics[trim={0 0 0 0},scale=0.34]{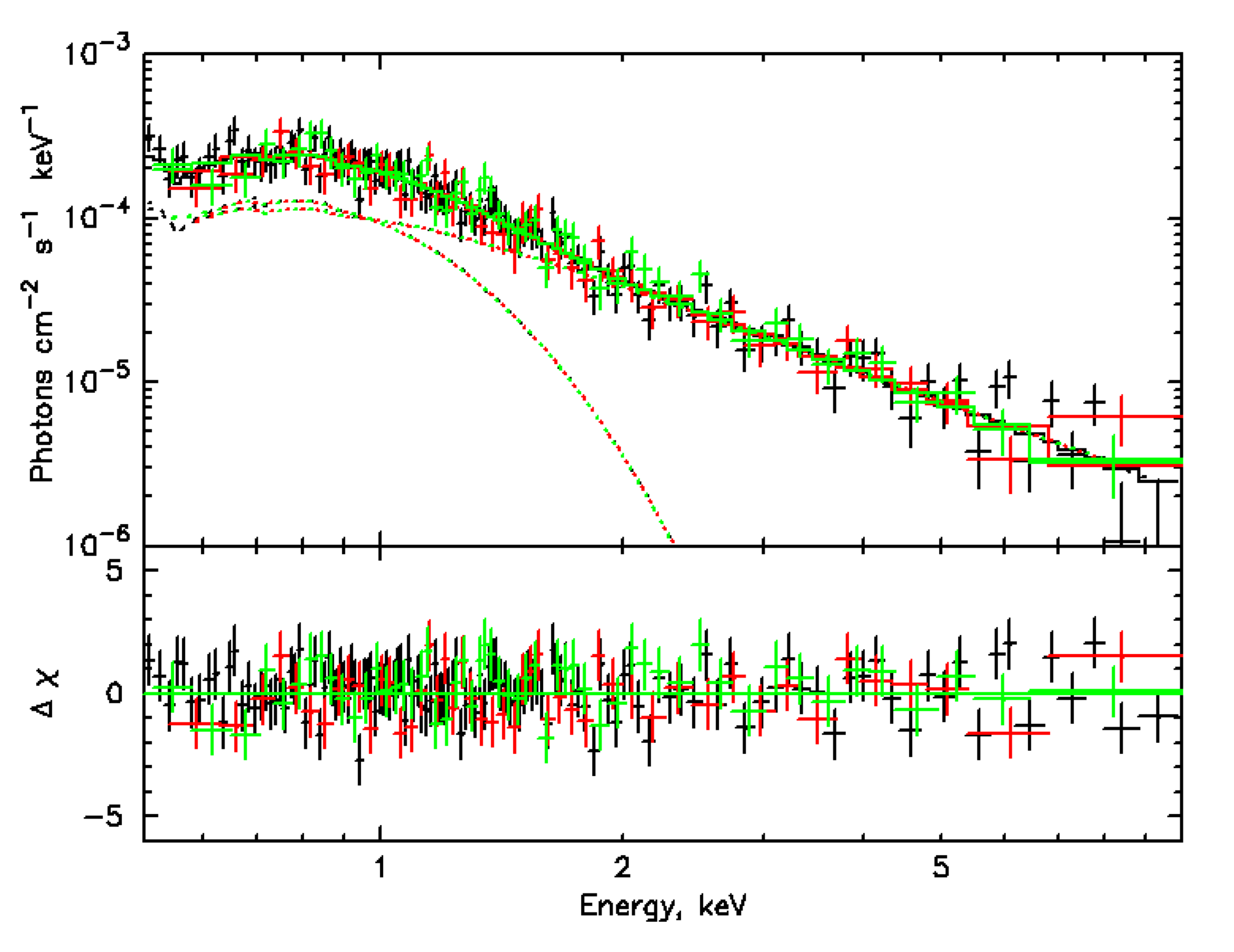}
\caption{ {\sl XMM-Newton} EPIC pn, MOS1, and MOS2 spectra of the brightest X-ray source in the field of  J1016 (see Section \ref{xspec}).  The best-fit BB+PL model is shown as dashed lines.
\label{xmm_spec}
}
\end{figure}

We have jointly fit the EPIC pn, MOS1, and MOS2 spectra of CXOU~J101546, having a combined total of 5650$\pm140$ net counts (the background contribution is $\sim8-11$\%).  We first fit an absorbed power-law (PL) model to the data. The best-fit model has an absorbing hydrogen column density $N_{\rm H}=(2.2\pm0.2)\times10^{21}$ cm$^{-2}$ and photon index $\Gamma=2.29^{+0.06}_{-0.05}$, with a reduced chi-squared $\chi^{2}_{\rm red}=1.17$ for $\nu=229$ degrees of freedom.  In order to further improve the fit, we added an absorbed blackbody (BB) component to the PL model. This model (shown in Figure \ref{xmm_spec}) has a best-fit absorbing hydrogen column density $N_{\rm H}=(2.4\pm0.4)\times10^{21}$ cm$^{-2}$, temperature $kT=0.20\pm0.02$ keV, BB  radius $R=430^{+170}_{-90}d_{1{\rm kpc}}$ m, and photon index $\Gamma=1.8\pm0.1$ with a reduced chi-squared $\chi^{2}_{\rm red}=0.98$ for $\nu=227$ degrees of freedom.

To evaluate the statistical significance of adding the additional BB component, we have used a likelihood ratio test\footnote{ The general procedure is outlined e.g., in Section 2.1 of \cite{2015PASA...32...38D} and \url{https://asd.gsfc.nasa.gov/XSPECwiki/statistical_methods_in_XSPEC}}, with 30,000 simulated spectra. We find the the BB+PL model is preferred over the PL only model at a $>4\sigma$ level, implying that the BB component is statistically necessary to adequately fit the spectra. {\sl XMM-Newton} lacks the angular resolution necessary to resolve the point source from the compact extended emission, and therefore, this spectrum is a combination of both components. The best-fit model gives an absorbed source flux of $F_X=(8.2\pm0.2)\times10^{-13}$ erg cm$^{-2}$ s$^{-1}$ and unabsorbed thermal and PL component fluxes of $2.4^{+0.4}_{-0.3}\times10^{-13}$ erg cm$^{-2}$ s$^{-1}$ and $(8.1\pm0.3)\times10^{-13}$ erg cm$^{-2}$ s$^{-1}$, respectively, in the 0.5-10 keV band.


 {\sl CXO} is able to resolve the point source  (having 33$\pm6$ net counts) from the surrounding extended emission  (having 57$\pm8$ net counts), but the short duration of the \cxo observation limits our ability to explore the spectra.
  Therefore, we chose to freeze $N_{\rm H}$ at the best-fit value obtained from the {\sl XMM-Newton} spectral fits to reduce the number of free parameters.
   For the point source, the best-fit absorbed BB model\footnote{The poor statistics do not allow us to test two-component models.} has $kT=0.35^{+0.06}_{-0.05}$ keV, $R=120^{+60}_{-30}d_{1{\rm kpc}}$ m,
    and an absorbed flux of $(1.6\pm0.3)\times10^{-13}$ erg  s$^{-1}$ cm$^{-2}$ (in 0.5-10 keV). The spectrum of the extended emission is well fit by an absorbed PL with  $\Gamma=1.7\pm0.3$ and absorbed flux, $F_{X}=(5.1^{+1.0}_{-0.9})\times10^{-13}$ erg  s$^{-1}$ cm$^{-2}$  (in  0.5-10 keV). The unabsorbed flux of the PL component is $(6.0^{+1.0}_{-0.9})\times10^{-13}$ erg  s$^{-1}$ cm$^{-2}$ (in 0.5-10 keV).   After fitting using W-stat, we calculate the Churazov weighted $\chi^{2}$ values to assess the goodness of fit \citep{1996ApJ...471..673C}. The weighted $\chi_{\nu}^{2}$ values are $\chi^{2}_{\nu}=1.01$ and $\chi^{2}_{\nu}=0.916$ for the point source and extended emission spectral fits, respectively, suggesting that the model fits the data reasonably well.

The spectrum of the large-scale extended emission seen in the {\sl XMM} observation was extracted from the white solid circle, south of the point source, shown in Figure \ref{xmm}.
 We only use the pn data because the extended emission partially overlaps with one of MOS 1's damaged CCDs, and MOS2 has fewer net counts than the pn.  Following the approach taken by, e.g., \cite{2016ApJ...824..138Y}, we first extracted the background spectrum from the region shown by the dashed white circle in Figure \ref{xmm}.  Then, we fit the background spectrum with a model containing a Bremsstrahlung component and a power-law component\footnote{\cite{2016ApJ...824..138Y} use two thermal components and two power-law components, however, our observation is substantially shorter so the background contains less counts and is well described by a simpler model.}. We exclude energies between 1.4 keV and 1.6 keV to avoid complications related to fitting the narrow Al K$\alpha$ instrumental line at $\sim1.49$ keV. We have checked that the exclusion of this line does not dramatically impact the resulting fit. We have also excluded energies above 7 keV because the extended emission becomes strongly background dominated. A total of 2056$\pm45$ counts remain in the background spectrum after these energy cuts are applied. For the purposes of our extended emission analysis, we are not concerned with the physical interpretation of the background model as long as it fits the background spectrum reasonably well.  The best-fit background model has a chi-squared $\chi^2=81$ for $\nu=73$ degrees of freedom and adequately describes the background spectrum.

 After fitting the background spectrum, the extended source spectrum (356$\pm46$ net counts with the same energy cuts as above) is fit with a combination of an absorbed PL model and the best-fit background spectral model. In this fit, the best-fit background model normalizations are scaled to account for  the differences in the region area and the exposure map between the background and source regions, and frozen to these scaled values. The $N_{\rm H}$ for the source PL model was frozen at the best-fit value for the point source (i.e., $N_{\rm H}=2.4\times10^{21}$ cm$^{-2}$). The best-fit  PL spectrum of the extended emission has $\Gamma=2.1\pm0.2$ and an unabsorbed flux $F=(1.6\pm0.2)\times10^{-13}$ erg cm$^{-2}$ s$^{-1}$ in the 0.5-10 keV energy range with a chi-squared $\chi^2=52$ for $\nu=55$ degrees of freedom. Following the same procedure, we have also fit the source spectrum with a bremsstrahlung model instead of a PL model, and find a best-fit temperature $kT=2.4^{+0.7}_{-0.5}$ keV and unabsorbed flux $F=(1.4\pm0.2)\times10^{-13}$ erg cm$^{-2}$ s$^{-1}$ in the 0.5-10 keV energy range with a chi-squared $\chi^2=48$ for $\nu=55$ degrees of freedom.

\subsubsection{Timing Analysis}
\label{xray_time}
We have searched for periodicity using the $Z_{n}^{2}$ (for n=1 and 2) test \citep{1983A&A...128..245B}. In the $10^{-4}-0.19$ Hz range we used the combined {\sl XMM-Newton} EPIC pn,  MOS1, and MOS2 barycenter corrected data extracted from a circle ($r=66\farcs5$ ) centered on CXOU~J101546, which contains 1,258, 1,329, 4,413 photons in MOS1, MOS2 and pn, respectively, in the 0.5-10 keV energy range. In addition, we used the pn data to search in the  0.19-6.8 Hz frequency  range afforded by the higher time  resolution of the pn detector.  We have also performed the search in 3 energy bands -- 0.5-10 keV, 0.5-2 keV, and 2-10 keV.  The most significant signal is  $Z_{1}^2=32.57$   in the energy range of 2-10 keV found at a frequency of  $5.871150(6)$ Hz using the pn-only data\footnote{This frequency is too high for the MOS detectors.}. This signal is significant only at the $2.6\sigma$ level\footnote{After taking into account the number of trial corresponding to 102054  independent frequencies where $T_{\rm span}$=15,008 s  is the time span of the EPIC pn exposure (see, e.g., \citealt{1999ApJ...511L..45P}  for details).  }, which is not  enough to be considered a  detection. Unfortunately, {\sl CXO} does not provide a high enough time resolution or number of counts to further test this periodicity.  Observations with higher time resolution with EPIC pn operated  in Small Window mode are needed to look for pulsations at higher frequencies typical for young pulsars. 

We have also searched for X-ray variability in the source during our {\sl XMM} observation by making a 500 s binned light curve. We find no evidence of variability in this light curve.


\subsection{Fermi-LAT $\gamma$-ray Data Analysis}
\label{lat_ana}

Here we describe the results of the spectral and timing analysis of J1016 using the {\sl Fermi-LAT} GeV data. 


\subsubsection{Spectral and Imaging Analysis}
\label{lat_ana_specim}

The 3FGL catalog\footnote{ The preliminary 4FGL catalog \citep{2019arXiv190210045T} was released while this paper was under review. We have compared our results to those reported in the catalog for 4FGL J1015.5-6030 (formerly 3FGL J1016.5--6034) and found them to be in good agreement.}, which is used to construct our model, was built from only four years of {\sl Fermi} data and was assembled using the Pass 7 IRF. {\sl Fermi} has now been observing the sky for $>10$ years, and the Pass 8 IRFs provide a number of dramatic improvements in the data \citep{2013arXiv1303.3514A}.  

We use a standard Binned Likelihood analysis\footnote{See \url{https://fermi.gsfc.nasa.gov/ssc/data/analysis/scitools/binned_likelihood_tutorial.html}}  with a spatial binning of 0.05$^{\circ}$and a spectral binning of four bins per decade of energy.
Our analysis begins by including a search for new sources that may have appeared due to the increased sensitivity of the current LAT data. We find three new sources within 3$^{\circ}$ of the 3FGL position of J1016 having a test-statistic (TS) larger than 40\footnote{The $TS$ is defined as $TS=-2\ln(L_{\rm 0}/L_{\rm 1})$ where $L_{\rm 0,max}$ is the maximum likelihood of the null hypothesis (i.e., model excluding the source) and L$_{\rm 1,max}$ is the maximum likelihood of the model including the source (see e.g., \url{https://fermi.gsfc.nasa.gov/ssc/data/analysis/documentation/Cicerone/Cicerone_Likelihood/Likelihood_overview.html}).}, the nearest of which is at a distance of $\sim1.2^{\circ}$. The new sources are all located within the Galactic plane ($b>-2^{\circ}$) and  could be caused by imperfections in the Galactic diffuse emission model \citep{2016ApJS..223...26A}. Therefore, we leave the detailed parameters of these new sources for the next FGL catalog.   
 For the remainder of our analysis, we include these sources in our model.

The LAT spectrum of J1016 was best fit by a LP spectral model in the 3FGL catalog \citep{2015ApJS..218...23A}  due to J1016's significant  spectral curvature (i.e., {\tt curvature\_significance}=5.8). However, we have carried out the spectral analysis steps described below using both a LP and PL model and find TS values of $\sim200$ and $\sim193$, respectively. Following the definition of the curvature significance used for the 3FGL catalog (see Section 3.3 in \citealt{2015ApJS..218...23A}),  we find that the source no longer has significant spectral curvature, possibly due to the additional exposure time, improvements in event reconstruction, and/or changes to the Galactic diffuse model. Therefore, we use the PL model for our  analysis.  
  To determine the position of  J1016 more accurately we restrict the photon energies to 0.3-300 GeV  (for the  source localization only) since the LAT PSF becomes  very broad at low energies and the source is relatively faint.  The updated position  (R.A.$=153.999^{\circ}$ and decl.$=-60.495^{\circ}$) and corresponding 2$\sigma$ error ellipse   ( $r_{\rm semi-major}=0.090^{\circ}$, $r_{\rm semi-minor}=0.069^{\circ}$, $\theta_{\rm ellipse}=172.58^{\circ}$) are shown in the left panel of Figure \ref{xmm}. Following the 3FGL catalog approach, we account for systematic uncertainties by multiplying both ellipse axes by a factor of 1.05 and adding an additional 0.005$^{\circ}$ in quadrature to each 95\% ellipse axis, leading to 95\% statistical plus systematic uncertainties of $r_{\rm semi-major}=0.095^{\circ}$, $r_{\rm semi-minor}=0.072^{\circ}$ \citep{2015ApJS..218...23A}. The error ellipse size has shrunk substantially compared to that from 3FGL and the position of CXOU J101546  is consistent with the $2\sigma$ positional uncertainty. For completeness, we note  that the LP fit gives a different position (R.A.=154.099, decl.=$-$60.487) that is offset by $\sim3'$  from the PL fit position, but the two error ellipses still overlap considerably and both of them include CXOU J101546 (see Figure \ref{xmm}).

\begin{figure}
\centering
\includegraphics[trim={0 0 0 0},scale=0.32]{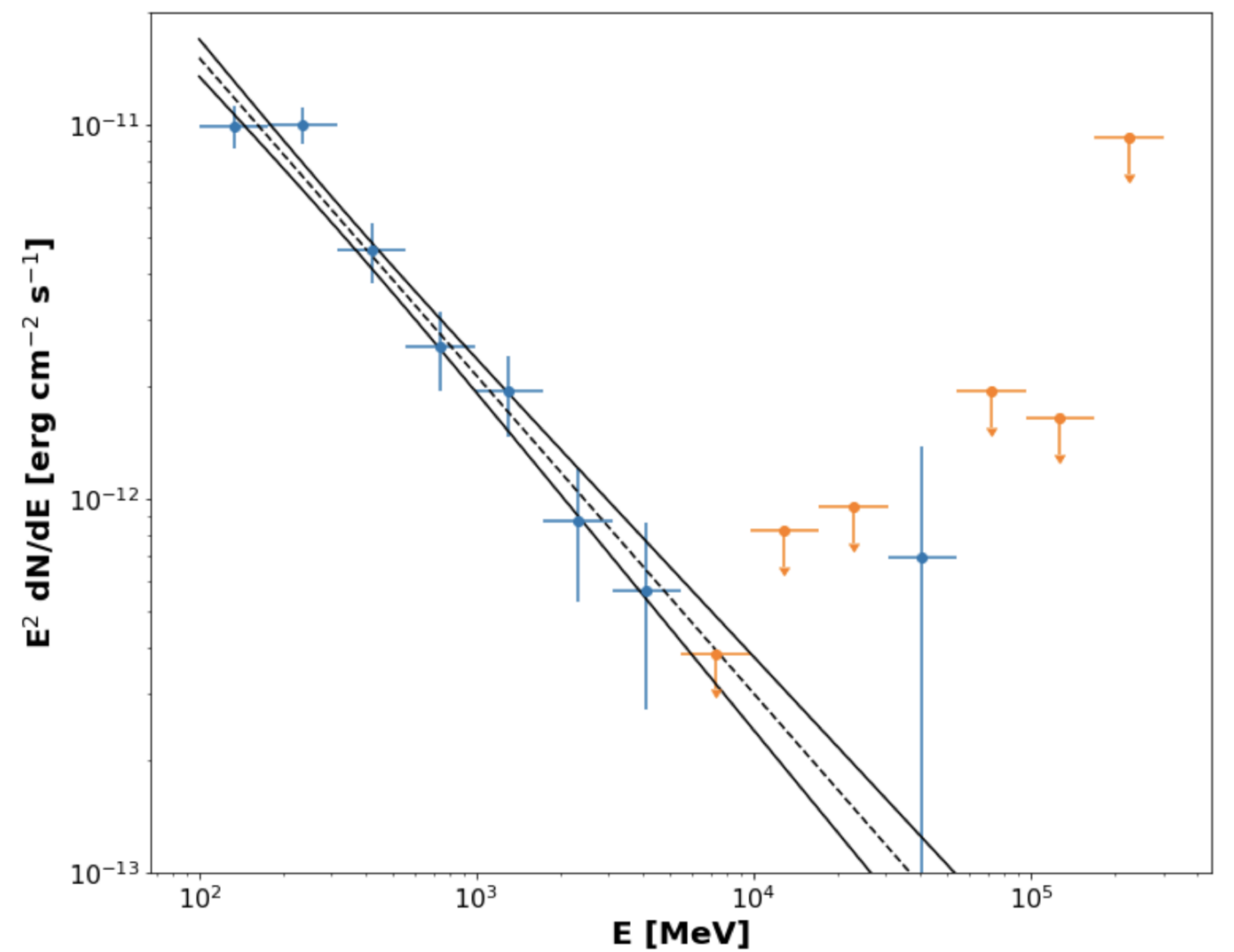}
\caption{ {\sl Fermi-}LAT GeV spectrum from 100 MeV to 300 GeV. The black dashed line represents the best-fit power-law model, while the solid black lines denote the $\pm1\sigma$ uncertainties. The blue points show the SED of 3FGL J1016 and the orange points show the 2$\sigma$ upper limits.
\label{gev_sed}
}
\end{figure}

Next, we fit the PL spectral model, fixing the source positions, while freeing  all components of the Galactic and isotropic diffuse emission, as well as the normalizations and spectral parameters of any point sources located within $3^{\circ}$ of the refined position of J1016. We then produced and inspected the TS and residual maps resulting from the best-fit PL model to ensure that there was no significant excess emission left in the vicinity of J1016. The best-fit PL model for J1016 has  $\Gamma=2.85\pm0.07$ and GeV flux $F=(1.8\pm0.1)\times10^{-11}$ erg  s$^{-1}$ cm$^{-2}$ (in 0.1-300 GeV)  with a $TS=192.73$. Lastly, we produce the spectral-energy distribution (SED) of J1016, which is calculated by fitting the flux normalization in a number of energy bins using a fixed-index PL spectral model (the photon index is fixed at the best-fit value $\Gamma=2.85$). The SED has a flux $F=(3.2\pm0.7)\times10^{-11}$ erg  s$^{-1}$ cm$^{-2}$ in  0.1-100 GeV  (see Figure \ref{gev_sed}). 


\subsubsection{Timing  Analysis}
\label{gev_time}
For the timing analysis we used the same filtering as described in Section  \ref{LATdata} but restricted the events to be within an  $r<0.7^{\circ}$ radius from the position of   CXOU~J101546.  We have also applied   a barycentric time correction to all  LAT events using the spacecraft orbit file with the  {\tt gtbary} tool.
 We then employed the approach described in  \cite{2014ApJ...795...75P} combined with the weighting scheme suggested by \cite{2018arXiv181206681B} and tested by \cite{2018arXiv181200719S}  to perform a blind search for pulsations in the  frequency and frequency derivative domain ($f=0-500$ Hz and $\dot{f}=-10^{-9}-0$ s s$^{-1}$). In the \cite{2014ApJ...795...75P}  algorithm we used a window size of $T_{\rm coh}=6$ days. Our search did not reveal any significant pulsations. We also checked the X-ray periodicity candidate frequency of  $5.871150(6)$ Hz using a modified   $Z_{n}^{2}$ (for n=1 and 2) including a modification to account for the frequency derivative but did not see any substantial periodic signal in the LAT data at  this frequency.
 
 To search for variability, we have also constructed the {\sl Fermi-}LAT light curves with weekly and monthly time bins. By fitting a straight line to the data, we find that the source is variable at the 1.5$\sigma$ level, which is not significant. Therefore, we conclude that the source is not variable at GeV energies.


\section{Discussion}
\label{sec4}
Although the $\gamma$-ray properties of J1016 are suggestive of having a pulsar origin, the $3^{\circ}$ offset from the Galactic plane is substantial, and an AGN association cannot be completely excluded based only on the GeV spectral properties and the lack of variability in the GeV light curve. X-ray observations of J1016, in combination with  multi-wavelength data at lower frequencies, shed additional light onto J1016's   nature  and  offer  strong support for the pulsar (and/or PWN) nature of J1016.

\begin{figure*}
\centering
\includegraphics[trim={0 0 0 0},scale=0.6]{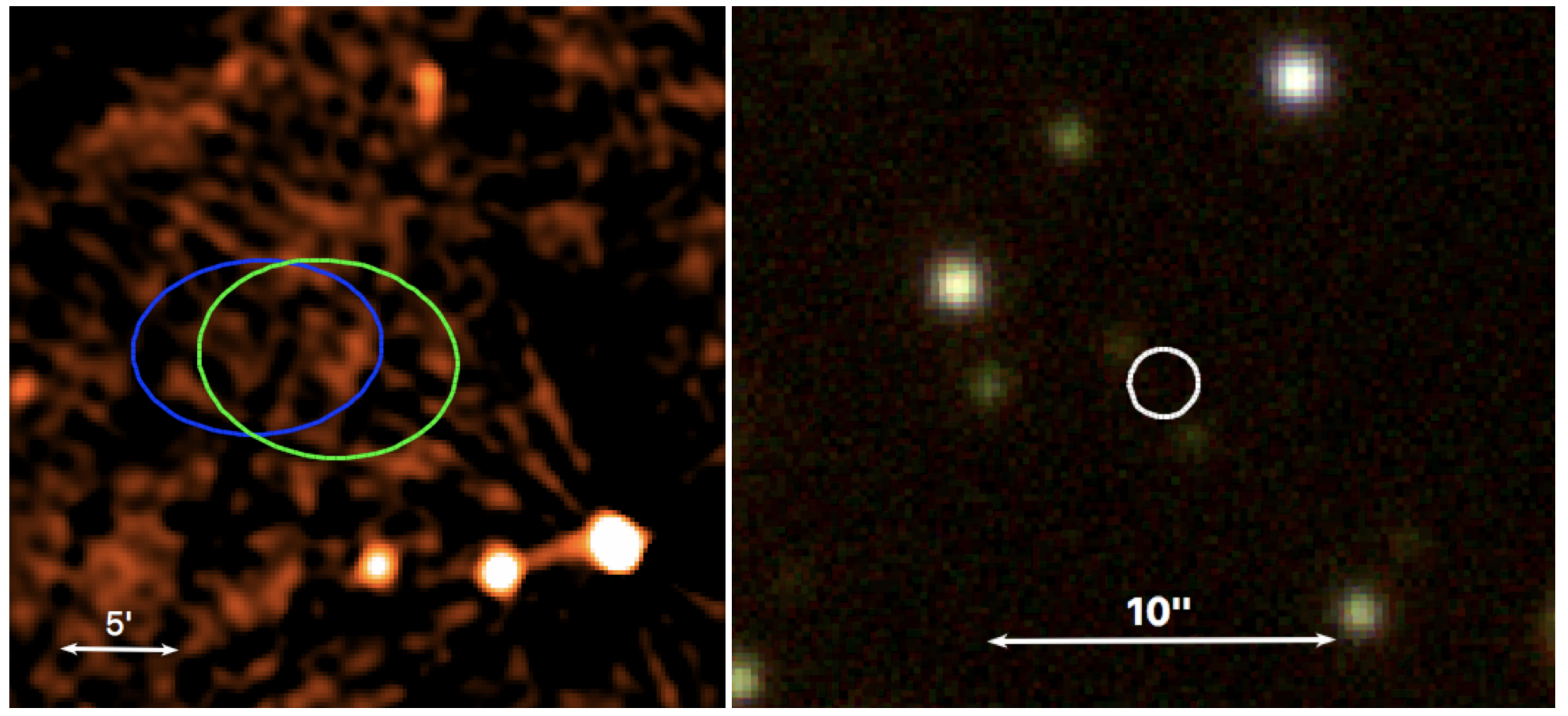}
\caption{The left panel shows the the  843 MHz radio image of the J1016 field from  the SUMSS  survey \citep{2003MNRAS.342.1117M}. The ellipses (green and blue for the PL and LP models, respectively) show the updated position and positional uncertainty using $\sim10$ years of {\sl Fermi-}LAT data. The right panel shown optical false-color (red: $i$-band, green: $r$-band, blue: $g$-band)  VPHAS+ image \citep{2014MNRAS.440.2036D} of the  CXOU~J101546 vicinity.  The  white circle ($r=0\farcs98$) is centered at the \cxo position of  CXOU~J101546 and shows the 2$\sigma$ positional uncertainty.  East is to the left and North is up in both images. 
\label{vphas}
}
\end{figure*}

The extended emission resolved on both  arcminute  (with {\sl XMM-Newton}) and arcsecond (with \cxo)
 scales is most naturally explained by a PWN powered by a young pulsar.  While some other interpretations could be considered for arcminute-scale emission, there is virtually no alternative for the compact emission which surrounds the point source CXOU J101546. This emission is too large, too bright (relative to the point source), and has too hard (i.e., $\Gamma=1.7$) of a spectrum for it to be a dust scattering halo. Although AGN can exhibit extended emission on arcsecond scales, it usually has an anisotropic morphology because it is associated with jets (see e.g., \citealt{2002ApJ...571..206S}). The  arcsecond scale emission around J1016 has a rather amorphous morphology which does not have any resemblance to that of jets. Lastly, galaxy clusters (GCs) filled with hot gas often appear as extended (on arcsecond to arcminute scales) objects in \cxo images (e.g., \citealt{2002astro.ph..7165F}). Some GCs host AGN which can produce GeV gamma-rays (e.g., Perseus A or IC 310 in the Perseus GC; \citealt{2011MNRAS.418.2154F,2017A&A...603A..25A}) and can also appear as relatively bright X-ray sources surrounded by diffuse X-ray emission from the GC.   However, GCs with AGN detected in GeV tend to be relatively nearby and, therefore, appear as bright acrminute-scale sources in X-ray, optical/IR, and radio images (see e.g., \citealt{2017A&A...603A..25A}). More importantly, AGN are typically variable sources in GeV and X-rays.  The X-ray emission surrounding   CXOU J101546 is faint and has a relatively small angular extent, there is no radio counterpart to J1016 in the SUMSS or VPHAS+ surveys (see Figure \ref{vphas}),   and there is no evidence of variability in GeV $\gamma$-rays or X-rays (see Sections \ref{xray_time} and \ref{gev_time}). Finally, the relatively low $N_{\rm H}$  value, compared to the Galactic $N_{\rm H}$ value, obtained for  CXOU J101546  from  the fit to the  {\sl XMM-Newton}  spectrum\footnote{Which, of course,  depends on the assumed spectral model.} argues against an extra-galactic origin (see  below). Therefore,  we consider the scenario of a GC hosting an AGN rather unlikely.

Another class of $\gamma$-ray sources that could, in principle, also exhibit extended X-ray emission with the observed morphology are high-mass $\gamma$-ray binaries (HMGBs; e.g., HMGBs B1259--63 \citealt{2014ApJ...784..124K,2015ApJ...806..192P} and possibly, LS 5039 \citealt{2011ApJ...735...58D}).  However, the optical/NIR image from the VPHAS+ survey (see Figure \ref{vphas}; \citealt{2014MNRAS.440.2036D})  does not reveal any counterpart to CXOU J101546, advocating against the HMGB scenario.  	Further, the nearest potential lower-wavelength counterpart lies 1\farcs6 away from the X-ray source, and is outside of the $\sim1\farcs5$ 3$\sigma$ positional uncertainty of CXOU J101546 \citep{2018A&A...616A...1G}. 

 Merely for completeness, we state that there is always the possibility that the bright extended X-ray source is coincident with J1016 by chance.  However, we consider this a very unlikely scenario because we are not aware of any such chance coincidences for other 3FGL sources. Additionally, the updated {\sl Fermi-}LAT position of J1016 is more consistent with the X-ray position of  CXOU J101546.0-602939 than in the 3FGL catalog, suggesting a true association.
 %

The spectrum of the point source fits a BB+PL model\footnote{Although the PL component must be at least partly attributed to the extended emission unresolved by {\sl XMM-Newton}. } which is very typical for  pulsars that are few tens of kyrs old (e.g., \citealt{2001ApJ...552L.129P,2007ApJ...660.1413K,2007ApJ...670..655K}). The spectrum of the compact nebula is well fit by  an absorbed  PL whose photon index, $\Gamma\approx1.7$,  is also typical for a PWN  (see \citealt{2008AIPC..983..171K}). The somewhat off-the-plane location of the source places a restriction onto the source's maximum distance, suggesting that it is unlikely to be larger than 5-6 kpc. On the other hand, it cannot be too much closer than a few kpc because then the BB radius would become too small for a young/middle-aged pulsar (unless we observe emission from just the polar cap, which is more common for pulsars with a few-hundred-year-old age).  At a fiducial distance of 3 kpc the luminosity of the point source, $L_{\rm psr}=5.7\times10^{31}$ erg s$^{-1}$, and compact PWN,  $L_{\rm pwn}=1.8\times10^{32}$ erg s$^{-1}$ in the 0.5-10 keV band,  which falls comfortably  in the range of PSR/PWN luminosities for pulsars with ages of 10-100 kyrs (see Figure 5 in \citealt{2008AIPC..983..171K} and Figure 2  in \citealt{2012ApJS..201...37K}). The best-fit hydrogen absorbing column density, $N_{\rm H}=2.4\times10^{21}$ cm$^{-2}$, of CXOU J101546  is a factor of $\sim3$ less than the total galactic  HI column density in this  direction, $\sim7\times10^{21}$ cm$^{-2}$ \citep{1990ARA&A..28..215D},  favoring a distance of a few kpc.  
At an adopted distance of 3 kpc, the $\gamma$-ray luminosity of J1016 would be $L_{\gamma}=1.1\times10^{34}d_{1}^2$ erg s$^{-1}$, which is also compatible with those of  pulsars detected at GeV energies (see Figure 9 in \citealt{2013ApJS..208...17A}).  The non-thermal X-ray flux of CXOU J101546 can be estimated by subtracting the  flux of the PWN (obtained from \cxo ACIS data) from the flux of the PL component in the fit to the {\sl XMM-Newton} EPIC data. This gives $\approx(2\pm1)\times10^{-13}$ erg  s$^{-1}$ cm$^{-2}$ in 0.5-10 keV.  Therefore, the GeV to non-thermal X-ray flux ratio of  CXOU J101546 ($F_{\rm 0.1-100 \ GeV}/F_{\rm 0.5-10 \ keV}\approx90-320$), is similar to those of some pulsars detected in X-rays and GeV $\gamma$-rays (see Figure 18 in \citealt{2013ApJS..208...17A}). 

  Assuming the large-scale extended emission is related to the X-ray point source, it could be also part of a PWN. For example, the 17 kyr-old pulsar, PSR J2021+3651, has both a  prominent compact PWN and rather luminous large-scale PWN \citep{2008ApJ...680.1417V}. Alternatively, it could be part of a SNR associated with CXOU J101546. However, the spectrum of appears to be somewhat too hard for that of a SNR.  On the other hand, it is possible that the large scale extended emission is unrelated to CXOU~J101546 and could belong to a background galaxy cluster that happens to be in the field-of-view. The spectrum of the large-scale extended emission is equally well fit by a bremsstrahlung model with a temperature of 2.4 keV, which is consistent with the measured temperatures of hot gas in galaxy clusters (see e.g., \citealt{2016ApJ...821...40S,2016A&A...595A..42T}). Deeper {\sl Chandra} observations are needed to resolve the large scale emission and to understand how it is related, if at all, to CXOU J101546.

If CXOU~J101546 is indeed a young pulsar then one could expect to see pulsations. However, we did not detect pulsations in X-rays or GeV $\gamma$-rays. The lack of X-ray pulsations is, however, not at all surprising. The time resolution of the Full Frame EPIC mode used in the observation of J1016 limits the pulsation search to periods longer than 144 ms for pn (and even larger periods for MOS), while many pulsars in 10-100 kyr range have shorter periods 
\citep{2013ApJS..208...17A}. However, the lack of reported $\gamma$-ray pulsations is more difficult to explain. The recent work of \cite{2018arXiv181200719S} shows that for some pulsars in the 10-100 kyr range, the detection of $\gamma$-ray pulsations can be very challenging  even when  the precise radio  ephemeris is available (e.g., PSR J1740+1000
 which has a X-ray PWN and PL+BB X-ray spectrum; \citealt{2008ApJ...684..542K,2012Sci...337..946K}).  Further, there are several examples of X-ray PWNe  which are X-ray and GeV $\gamma$-ray sources, where neither X-ray nor $\gamma$-ray pulsations have been found (e.g., PWNe in  SNR G327.1-1.1\footnote{This source is currently listed as a GeV source in the preliminary version of the 8-yr {\sl Fermi}-LAT catalog (see \url{https://fermi.gsfc.nasa.gov/ssc/data/access/lat/fl8y/}).} and MSH 11-62;  \citealt{2009ApJ...691..895T} and \citealt{2012ApJ...749..131S}, respectively). 

\section{Summary and Conclusions}
\label{sec5}

We have discovered a new X-ray source, likely a young pulsar with a PWN, powering the unidentified {\sl Fermi-}LAT source 3FGL J1016. The {\sl XMM} point-source spectrum is well fit by an absorbed blackbody+power-law model, with a temperature $kT=0.2\pm0.02$  keV and photon index $\Gamma=1.8\pm0.1$, which is typical of young-to-middle-aged pulsars. Both the compact nebula resolved by {\sl CXO} and the large-scale extended emission observed with {\sl XMM} can be fit with a power-law, with photon indices of $\Gamma=1.7$ and 2.1, respectively. The pulsar is likely to be at a distance of a few kpc. No periodicity has been detected at X-ray or GeV energies. Deeper X-ray observations with \xmm EPIC-pn in Small Window mode  and dedicated  searches in radio (and $\gamma$-rays),  taking advantage of the precise \cxo position, may detect pulsations in the future.

\medskip\noindent{ Acknowledgments:}
We thank the anonymous referee for their helpful and constructive feedback which has helped to improve the manuscript. Support for this work was provided by the National Aeronautics and Space 
Administration through the awards 80NSSC18K0635 and 80NSSC17K0760 and  
the Chandra award GO8-19063X issued
by the Chandra X-ray Observatory Center, which is operated by the 
Smithsonian Astrophysical Observatory for and on behalf of the National 
Aeronautics Space Administration under contract
NAS8-03060. Based on data products from observations made with ESO Telescopes at the La Silla Paranal Observatory under program ID 177.D-3023, as part of the VST Photometric H$\alpha$ Survey of the Southern Galactic Plane and Bulge (VPHAS+, www.vphas.eu). This work has made use of data from the European Space Agency (ESA) mission
{\it Gaia} (\url{https://www.cosmos.esa.int/gaia}), processed by the {\it Gaia}
Data Processing and Analysis Consortium (DPAC,
\url{https://www.cosmos.esa.int/web/gaia/dpac/consortium}). Funding for the DPAC
has been provided by national institutions, in particular the institutions
participating in the {\it Gaia} Multilateral Agreement.

\end{document}